\documentclass[12pt,a4paper,english]{iopart}
\pdfoutput=1
\usepackage[T1]{fontenc}
\usepackage[latin1]{inputenc}
\usepackage{graphicx}
\usepackage{amssymb}

\makeatletter

\providecommand{\tabularnewline}{\\}

\usepackage{setstack}

\usepackage{cmap}

\makeatother

\usepackage{babel}

\begin{document}

\title[Benzene, phenol, propane and carbonic acid on oxidized aluminium surfaces]{Adsorption of benzene, phenol, propane and carbonic acid molecules on oxidized Al(111) and $\alpha$-Al$_{2}$O$_{3}$(0001) surfaces: A first-principles study}

\author{Janne Blomqvist and Petri Salo}

\address{Department of Applied Physics, Helsinki University of Technology,
P.O. Box 1100, FI-02015 TKK, Finland}

\begin{abstract}
We present the results of \emph{ab initio} calculations describing
the adsorption of certain small organic molecules on clean and oxidized
Al(111) surfaces as well as on the $\alpha$-Al$_{2}$O$_{3}$(0001)
surface. Our results show that adsorption of benzene on the clean
and oxidized Al(111) surfaces is generally weak, the adsorption energy
being at most around $-0.5$ eV per benzene molecule, and the molecule
adsorbed at a considerable distance from the surfaces. The adsorption
energy varies weakly at the different adsorption sites and as a function
of the oxygen coverage. For the alumina surface, we find no benzene
adsorption at all. We have also calculated a phenol molecule on the
aluminium and alumina surfaces, since it is similar to the benzene
molecule. The results show a weak adsorption for phenol on the alumina
surface and no adsorption on the aluminium or oxidized aluminium surfaces
at all. For the propane molecule there is no adsorption on either
the oxidized aluminium or the alumina surface, whereas the carbonic
acid molecule binds strongly to the alumina but not to the aluminium
surface.
\end{abstract}
\maketitle
Aluminium, Oxygen, Aromatics, Carbonic acid, Propane, Density functional
calculations, Adsorption

\section{Introduction}

Hybrid materials, where multiple substances are combined, belong to
an interesting new class of materials. These materials can be optimized
for specific applications, emphasizing appearance, strength or some
other property \cite{straznicky2004}. To understand these materials
we must understand the interface between the various substances. 

Particularly interesting materials from a practical as well as a scientific
perspective are the materials that combine a metal and polymer. Studying
the interface between these substances at the microscopic scale is
of a great importance. Thus, studying the interaction between the
metal surface and the molecules in the polymer chains is important.
There are also multiple adhesion mechanisms that need to be understood,
such as mechanical gripping, and multiple phenomena affecting the
strength of the interface, such as the surface roughness and wetting.
In this work, however, we have focused on the microscopic scale quantum
mechanical binding between the aluminium surface and organic compounds
common in polymers using first-principles calculations.

Real aluminium surfaces oxidize very quickly when in contact with
the atmosphere. To investigate how the oxygen coverage affects adsorption,
we have used density functional theory to calculate the adsorption
energy at different oxygen coverages. Aluminium is also often electrolyzed
to increase corrosion resistance. During the electrolyzing process,
a layer of alumina is formed on the aluminium surface. Thus, we have
also investigated the adsorption of the molecules on the $\alpha$-alumina
(0001) surface. The (0001) surface of alumina was chosen because it
has been extensively studied in the past, both experimentally and
computationally. 

Previously there has been relatively little work on the systems of
organic molecules on aluminium surface, either computational or experimental.
Duschek et al. \cite{duschek2000} studied benzene on clean Al(111)
both computationally and experimentally, and Chakarova-Käck et al.
\cite{chakarova2006} studied phenol on $\alpha$-Al$_{2}$O$_{3}$(0001)
using density functional theory calculations with van-der-Waals corrections.
Su et al. \cite{su1997} studied carbonic acid anions in solution
adsorbing on an amorphous alumina surface, showing that carbonic acid
adsorbs on top of the Al atom in the alumina substrate. Similarly,
Alliot et al. \cite{alliot2005} studied carbonic acid anions on an
$\alpha$-Al$_{2}$O$_{3}$ surface, showing the presence of adsorption.
Johnston et al. \cite{johnston2007} studied the adsorption of the
bisphenol-A-polycarbonate (BPA-PC) on the Si(001) surface component
molecules using density functional theory calculations.

In this paper we have calculated the adsorption of benzene, phenol,
carbonic acid and propane on the close packed Al(111) and $\alpha$-Al$_{2}$O$_{3}$(0001)
surfaces. With these molecules one could build up a BPA-PC molecule,
which is an important polycarbonate plastic. The benzene molecule
has been placed at the fcc, hcp, bridge and top sites, with different
amounts of oxygen on the Al(111) surface. The oxygen molecule is either
at the fcc or hcp hollow site, or there are two nearest neighboring
oxygens at adjacent fcc and hcp sites. Finally, we have also calculated
the case where the Al(111) surface is covered by one monolayer (ML)
oxygen at the fcc sites. For the other components and the $\alpha$-Al$_{2}$O$_{3}$(0001)
surface only a few configurations have been calculated, with the component
molecule on the top of an Al and O atom or at the hollow sites. 

The rest of the paper is organized as follows. First, in Section \ref{sec:Methods}
we give a brief outline of the computational techniques we have used,
in Sections \ref{sub:Al(111)-surface} and \ref{sub:Al_{2}O_{3}-(0001)-surface}
we present our results for the Al(111) and $\alpha$-Al$_{2}$O$_{3}$(0001)
surfaces, respectively, and finally we discuss our results and give
some concluding remarks in Section \ref{sec:Discussion}.

\section{\label{sec:Methods}Methods}

We have used the \emph{Vienna Ab Initio Simulation Package} (VASP)
\cite{kresse1993,kresse1996a,kresse1996b}, a program for calculating
the energy ground state using a plane wave basis set. The many-body
effects have been taken into account with the revised version \cite{hammer1999}
of the Perdew-Burke-Ernzerhof (PBE) \cite{perdew1996} form of the
generalized gradient approximation\emph{,} which is known to produce
better molecular adsorption energies compared to vanilla PBE, however,
at the expense of a bigger error in the lattice constant compared
to full potential calculations\emph{.} For the potentials we have
used the projector augmented wave method \cite{blochl1994,kresse1999}
with a plane wave cutoff energy of 400 eV. 

For the Al(111) calculations, a supercell with 4x4x4 Al atoms has
been used, while for the alumina calculations we have used a larger
Al-terminated supercell with 48 Al and 72 O atoms in 12 Al and 6 O
layers, forming a 2x2 hexagonal surface supercell. The thickness of
the slab for the Al case was 7.9 Å, while the alumina slab was thicker
being 12.2 Å, in order to take into account the large surface relaxations
in the Al-terminated alumina surface \cite{godin1994}. For the Al
case, there was 16.5 Å vacuum between the periodic slabs, while for
alumina the corresponding value was 14.0 Å. 

The Brillouin zone was sampled using the Monkhorst-Pack scheme \cite{monkhorst1976,pack1977},
and the number of k points in the irreducible part of the Brillouin
zone was 15 (5x6x1 mesh) for the oxygen covered Al(111) surface and
4 k-points (2x2x1 $\Gamma$-centered mesh) for the $\alpha$-Al$_{2}$O$_{3}$(0001)
surface. In the case of the Al(111) surface the k-point grid was nonuniform
in the surface plane since the surface cell used was rectangular,
and thus the length of the lattice vectors in the plane were not equal.
Hence we tested different combinations of k-point grids before choosing
the 5x6x1 grid. To make the convergence towards the electronic ground
state faster the Fermi level was smeared using the method of Ref.
\cite{methfessel1989} with a value of 0.2 eV for the metal surface,
and for the alumina surface we used Gaussian smearing with a smearing
parameter of 0.4 eV. Test calculations were made with spin polarization
enabled; no spin polarization was seen and thus the rest of the calculations
were made without spin polarization to reduce the computational burden.
In order to account for potential charge inbalance on the surfaces
in the cell, a dipole correction was applied in the direction normal
to the surface.

Atomic coordinates were relaxed until forces were less than 0.02 eV/Å,
and k-point grids were chosen to make the error in total energy less
than 0.05 eV. The same procedure was followed during the adsorption
energy calculations, that is, the molecules were placed in approximate
starting positions and allowed to relax freely until forces were less
than 0.02 eV/Å. 

For the analysis of the local density-of-states (LDOS), one needs
to compare the LDOS's of different systems with each other. Thus,
the energy scales have to be comparable. For the reference system
for each plot, typically the surface slab with a molecule adsorbed,
the calculated Fermi energy $E_{F}$ was subtracted, \[
E'=E-E_{F}\,,\]
where $E$ is the tabulated energy. For the calculations used for
comparing with the reference system, such as a molecule in vacuum,
the energy was shifted as\[
E'=E-E_{F}-(\phi-\phi_{ref})=E-E_{F,ref}-(E_{vac}-E_{vac,ref})\,,\]
where $\phi=E_{vac}-E_{F}$ is the work function of the system, $\phi_{ref}$
is the work function of the reference system, $E_{vac}$ is the maximum
\emph{z}-averaged vacuum potential in the supercell of the system,
and $E_{vac,ref}$ is the maximum \emph{z}-averaged vacuum potential
for the reference system.

\section{Results and discussion}

In this section we show our results for the molecules on the Al(111)
and $\alpha$-Al$_{2}$O$_{3}$(0001) surface. The molecules with
their chemical compositions are benzene (C$_{6}$H$_{6}$), phenol
(C$_{6}$H$_{5}$OH), carbonic acid (H$_{2}$CO$_{3}$), and propane
(C$_{3}$H$_{8}$). We define the coverage of a molecule on the surface
as the number of molecules per atoms in the surface layer. For each
system we calculate the electronic and geometric structure and the
total energy. We define the adsorption energy as\[
E_{ads}=E_{tot}-E_{slab}-E_{mol}\,,\]
where $E_{tot}$ is the total energy of the system with the molecule
or atom adsorbed on the surface, and $E_{slab}$ is the total energy
of the empty surface slab. The third term, $E_{mol}$ is the total
energy of the adsorbed molecule or atom in vacuum calculated using
the same cell geometry, k-point grid and other calculation parameters
such as smearing, as with the slab calculations. In the case of the
calculations on the oxidized surface, $E_{slab}$ is the total energy
of the oxidized surface slab.

\subsection{\label{sub:Al(111)-surface}Al(111) surface}

With the parameters described in the preceding section, the equilibrium
lattice constant for bulk fcc aluminium has been calculated to be
$4.055$ Å, compared to the experimental result of $4.0496$ Å \cite{CRC86}.
For comparison, with Perdew-Wang 91 \cite{perdew1992} and the regular
PBE gradient approximations the lattice constants are found to be
4.05 and 4.04 Å, respectively \cite{mattsson2006}, and using the
PBE approximation 4.052 Å \cite{duschek2000}.

For the Al(111) surface with an oxygen coverage, the adsorption energy
per oxygen atom has been calculated, and the results can be seen in
Table \ref{cap:O-Adsorption-energy}. As expected, the highest adsorption
energy for a single O atom was found at the fcc site. When the entire
surface was covered with 1 ML oxygen at the fcc sites, the adsorption
energy per oxygen atom was $0.44$ eV lower than for an isolated oxygen
atom, indicating that the aluminium surface oxidizes very easily.
For the fcc and hcp sites, the results are comparable to other calculations
\cite{yourdshahayan2002}. More detailed calculations describing the
behavior of oxygen on Al(111) can be found in Refs. \cite{yourdshahayan2002,kiejna2001,kiejna2002}.

\begin{table}
\caption{\label{cap:O-Adsorption-energy}Adsorption energy for O on Al(111)
per O atom at different coverages. Individual atoms at the fcc and
hcp sites, two oxygen atoms at nearest neighbor (nn) fcc and hcp sites,
and 1 ML of oxygen atoms at the fcc sites. The energies are with respect
to atomic oxygen in vacuum; with respect to an half of the energy
of the O$_{2}$ molecule in vacuum, $3.19$ eV should be added to
the tabulated values.}

\begin{centering}
\begin{tabular}{c|c}
 & Ads. energy (eV)\tabularnewline
\hline
fcc/0.0625 ML & $-7.27$\tabularnewline
hcp/0.0625 ML & $-6.94$\tabularnewline
nn/0.125 ML & $-7.00$\tabularnewline
fcc/1 ML & $-7.71$\tabularnewline
\hline
\end{tabular}
\par\end{centering}
\end{table}

\subsubsection{\label{sub:Benzene}Benzene}

\begin{figure}
\begin{centering}
\includegraphics[width=0.35\columnwidth]{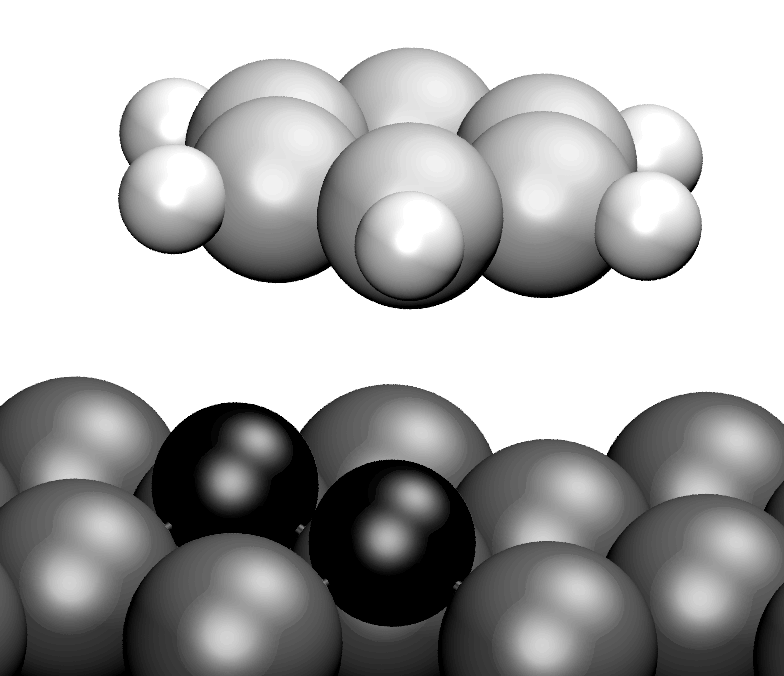}\includegraphics[width=0.35\columnwidth]{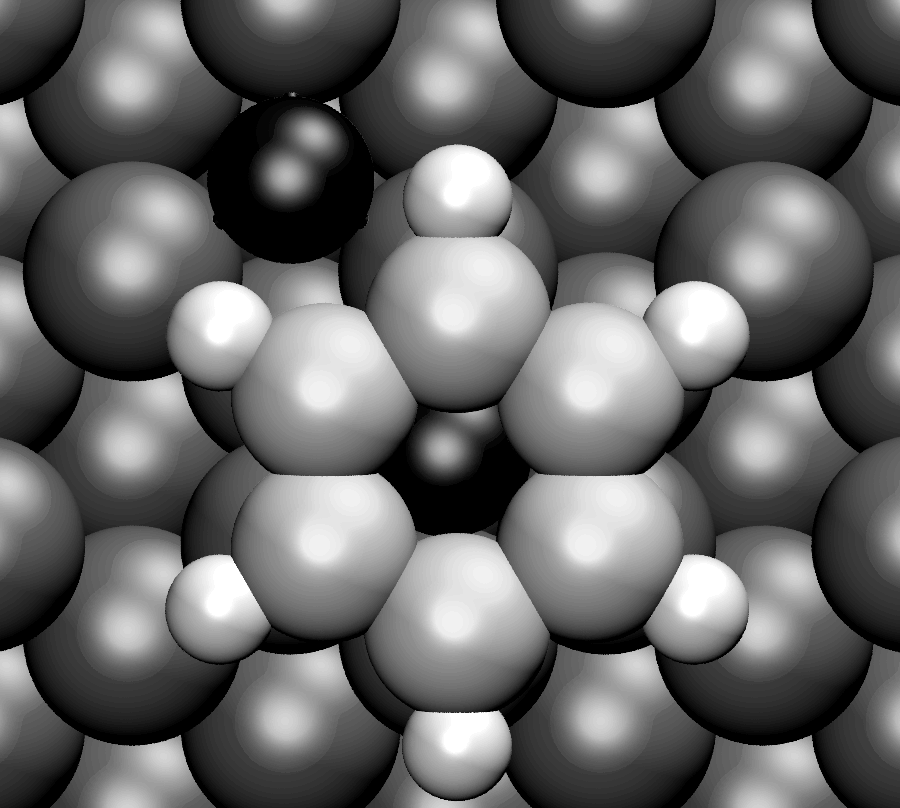}
\par\end{centering}

\caption{\label{cap:al111-benz-fcc}A side (left) and top (right) view of the
benzene molecule on top of the fcc site on the Al(111) surface, without
the $\frac{\pi}{6}$ rotation and with nearest neighbor oxygens on
the surface. Hydrogen atoms are seen as small light gray spheres,
carbon as large light gray spheres, oxygen as large black spheres,
and aluminium as large dark gray spheres.}

\end{figure}

In Table \ref{cap:Benzene-on-Al(111)} one can see the adsorption
energy of benzene as it adsorbs on the Al(111) surface with different
O coverages. One can see that with the exception of the configuration
with nearest neighbor oxygens at the fcc and hcp sites, where the
adsorption energy is about $-0.5$ eV, there is very weak or no adsorption
at all. In Fig. \ref{cap:al111-benz-fcc}, one can see an example
of a configuration, with the benzene molecule at the fcc site. 

\begin{table*}
\caption{\label{cap:Benzene-on-Al(111)}Adsorption energy (eV) for benzene
at different sites (with and without $\frac{\pi}{6}$ rotation around
the axis perpendicular to the surface) on Al(111) with the benzene
coverage of $0.063$ ML and with different oxygen coverages.}

\begin{centering}
\begin{tabular}{c|cccccccc}
 O$\diagdown$C$_{6}$H$_{6}$ &  fcc  &  fcc $\frac{\pi}{6}$ &  hcp  &  hcp $\frac{\pi}{6}$ &  bridge  &  bridge $\frac{\pi}{6}$ &  top  &  top $\frac{\pi}{6}$\tabularnewline
\hline
clean  &  $0.12$  &  $0.12$  &  $0.08$  &  $0.12$  &  $0.12$  & $0.12$  &  $0.21$  &  $0.21$ \tabularnewline
fcc  &  $0.16$  &  $0.16$  &  $0.25$  &  $0.25$  &  $0.15$  &  $0.16$  &  $0.14$  &  $0.15$ \tabularnewline
hcp  &  $0.19$  &  $0.20$  &  $0.33$  &  $0.32$  &  $0.20$  &  $0.20$  &  $0.20$  &  $0.19$ \tabularnewline
nn  &  $-0.52$  &  $-0.51$  &  $0.13$  &  $0.14$  &  $-0.52$  &  $-0.52$  &  $-0.52$  &  $-0.51$ \tabularnewline
1 ML &  $0.17$  &  $0.17$  &  $0.17$  &  $0.18$  &  $0.17$  &  $0.18$  &  $0.17$  &  $0.18$ \tabularnewline
\hline
\end{tabular}
\par\end{centering}
\end{table*}

The work functions for the surfaces were also calculated. In all the
cases except for the one with 1 ML oxygen, the work function was between
4.0 and 4.1 eV. For the 1 ML case, it was between 4.5-5.0 eV depending
on the location of the benzene molecule.

\begin{figure}
\begin{centering}
\includegraphics[width=0.7\columnwidth]{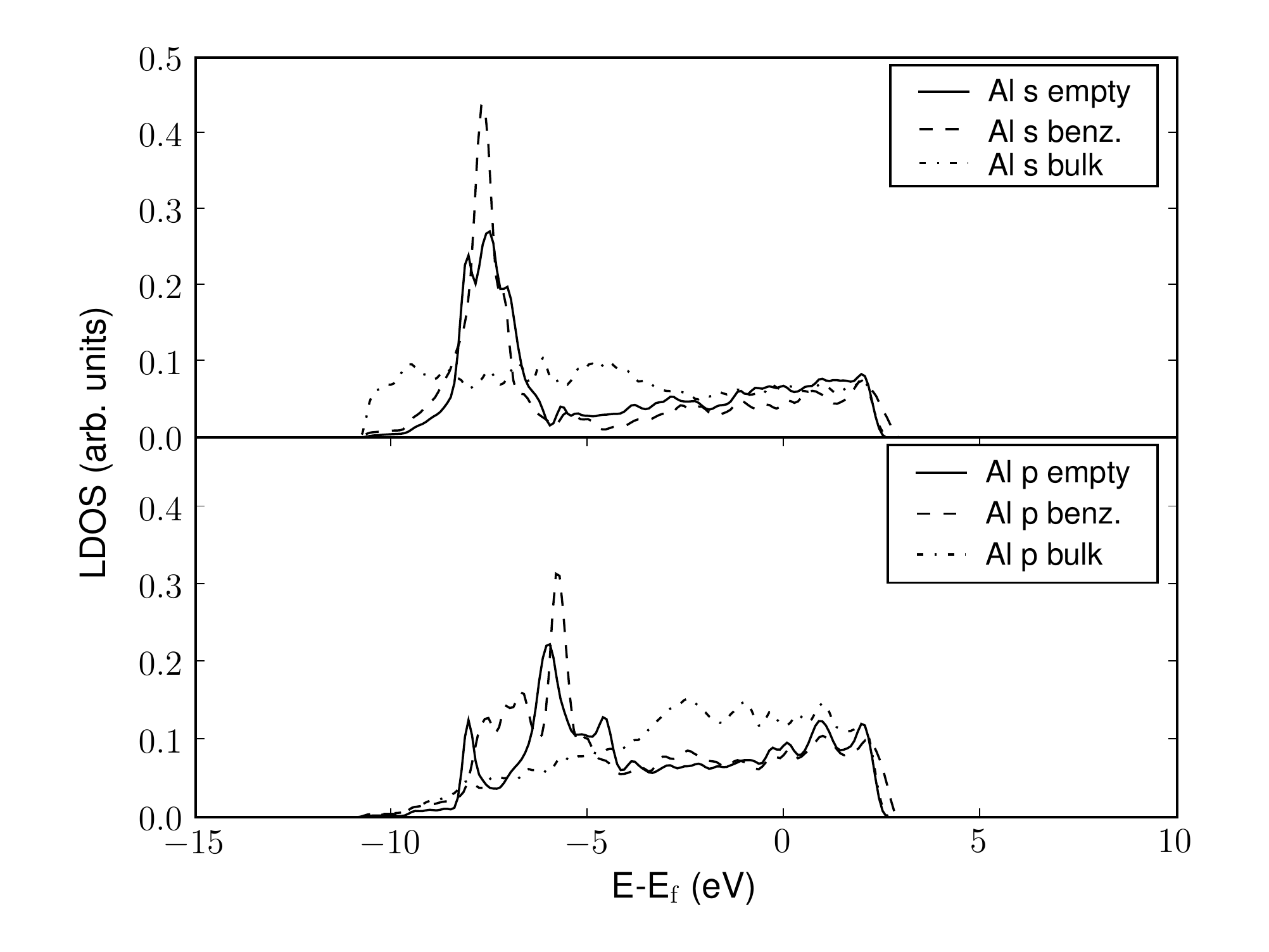}
\par\end{centering}

\caption{\label{cap:Al111-ldos}LDOS for the Al atoms in the surface layer
of the Al(111) surface with nearest neighbor oxygens present, without
(solid lines) and with (dashed lines) a benzene molecule (fcc $\frac{\pi}{6}$
site). For comparison the LDOS for an atom in bulk Al (chain line)
can also be seen. Coverage of C$_{6}$H$_{6}$ is 0.063 ML.}

\end{figure}

\begin{figure}
\begin{centering}
\includegraphics[width=0.7\columnwidth]{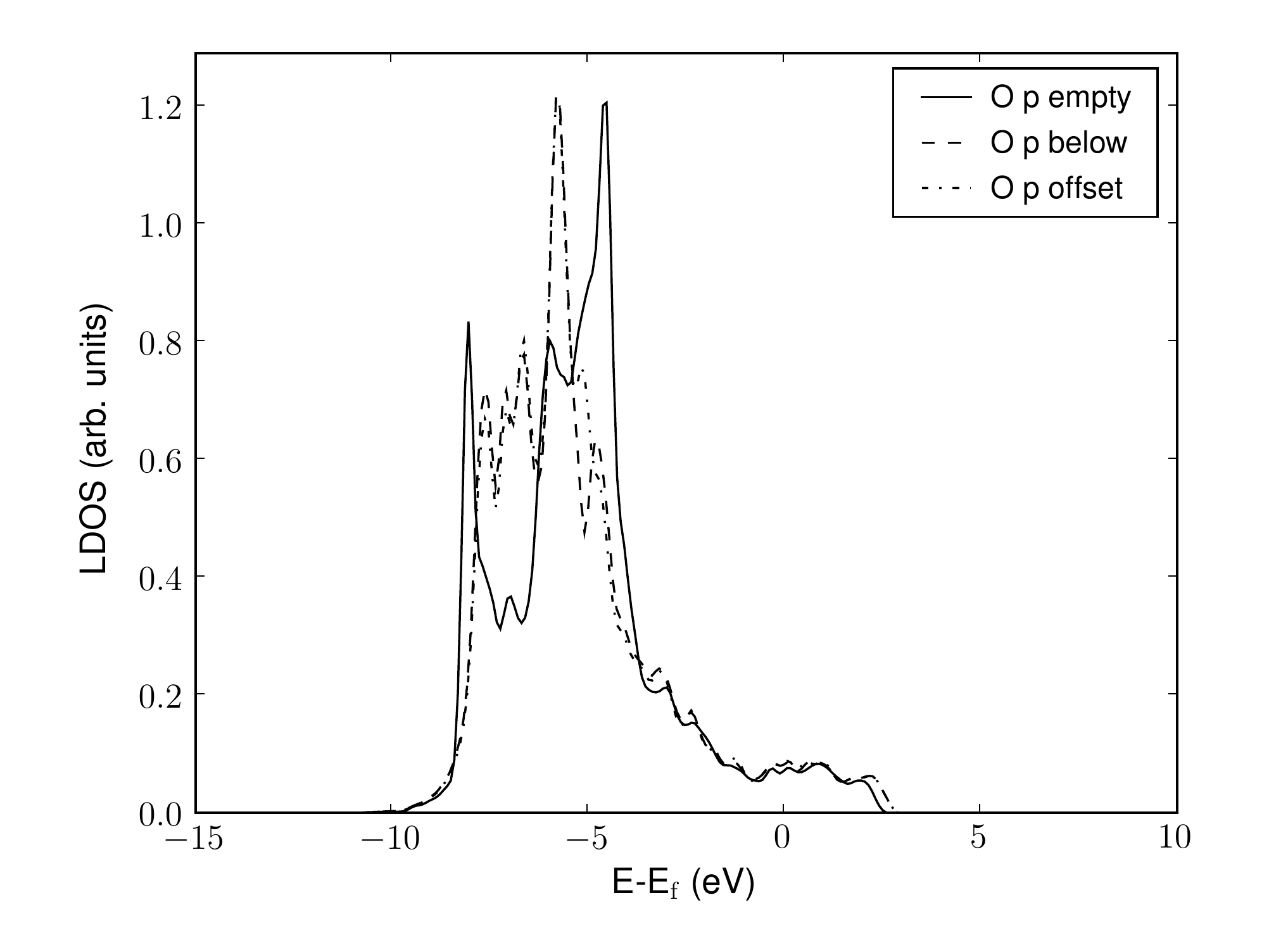}
\par\end{centering}

\caption{\label{fig:Al111-nn-O-dos}p-LDOS for the nearest neighbor oxygens
on the Al(111) surface, without (solid line) and with (dashed and
chain lines) a benzene molecule at the fcc $\frac{\pi}{6}$ site.
When the benzene molecule is present, the oxygen is either directly
below the center of the benzene molecule (dashed line), or offset
from the center by about 2.4 Å in the next hollow site (chain line).
Coverage of C$_{6}$H$_{6}$ is 0.063 ML.}

\end{figure}

In all the cases, the equilibrium distance above the surface for the
benzene molecule was about 4.0 Å, also for the cases of the nearest
neighbor O on Al(111) where the adsorption of benzene was observed.
By comparing charge density differences between the benzene molecule
and Al(111) slab separately and when the molecule was on the surface,
we saw that there was no charge transfer between the benzene molecule
and the surface. To analyze further the situation we have plotted
the LDOS with specific angular projections for different atoms in
the system. In Fig. \ref{cap:Al111-ldos}, for Al atoms in the surface
layer on the Al(111) surface with nearest neighbor oxygens on the
surface, one can see that there is practically no difference in the
LDOS of Al for the cases where the Al atom is below the benzene or
on an empty surface. However, in Fig. \ref{fig:Al111-nn-O-dos} one
can see that the presence of the benzene molecule causes a change
in the p-LDOS of the oxygen atoms on the surface. There is almost
no difference whether the oxygen atom is directly below the center
of the benzene molecule or slightly offset. Thus, the adsorption of
benzene has to be due to the mutual binding to both of the oxygen
atoms on the surface (see also Table \ref{cap:Benzene-on-Al(111)}).

We can compare our results with other experimental and computational
results \cite{duschek2000}. In contrast to the value reported for
the equivalent clean surface configuration in Ref. \cite{duschek2000},
$-0.351$ eV, we found no adsorption on the clean surface, though
it should be mentioned that we used a different potential functional.
Despite our efforts, we have been unable to reproduce the adsorption
energy reported there, even if we used the same potential and with
a denser k-point grid than normal (11x11x1). However, the other observations
in Ref. \cite{duschek2000}, that i.e. the density of states are not
changed substantially and the distance of C$_{6}$H$_{6}$ from the
surface is quite large, as well as the photoemission and thermal desorption
spectras, indicate that the binding of C$_{6}$H$_{6}$ to the Al(111)
surface is rather weak, which is in line with our results.

We have also calculated the C$_{6}$H$_{6}$ coverage dependence on
the adsorption energy. We only checked the system of benzene at the
fcc site with the nearest neighbor oxygen on the Al(111) surface.
By putting two benzene molecules in the original 4x4 supercell, the
adsorption energy for the coverage of 0.125 ML was found to be $-0.51$
eV which does not differ too much from the value of $-0.52$ eV for
the 0.063 ML case. Thus, we saw no coverage dependency in the adsorption
energy.

\subsubsection{Phenol}

As the adsorption energy of the benzene molecule was not particularly
dependent on the adsorption site, we decided not to test the phenol
molecule (C$_{6}$H$_{5}$OH) as exhaustively as the benzene molecule.
In all cases the phenol molecule was placed over the surface such
that the oxygen atom in the phenol was on the top of the fcc site.
Thus, as we previously found out that benzene does not adsorb to the
clean surface, and when it shows some adsorption there must be some
oxygen on the surface, the working hypothesis is that the phenol molecule
would bind to the surface via the oxygen atom, with the hydrogen atom
in the hydroxyl group potentially dissociating. The results can be
seen in Table \ref{tab:al-phenol}. In general, the results are comparable
to those of benzene, i.e. there is no adsorption at all. The main
difference is that while for the case of benzene, the nearest neighbor
configuration (where there were two oxygen atoms in neighboring fcc
and hcp sites) exhibited some adsorption, for phenol there is no adsorption
in that case either.

\begin{table*}
\caption{\label{tab:al-phenol}Adsorption energies of phenol on the clean and
oxidized Al(111) surface. In all cases the oxygen in the phenol molecule
is at the fcc hollow site on the (111) surface.}

\begin{centering}
\begin{tabular}{c|ccccc}
\multicolumn{1}{c|}{} & clean & fcc & hcp & nn & 1 ML\tabularnewline
\hline
Ads. energy (eV) & $0.07$ & $0.08$ & $0.07$ & $0.02$ & $0.00$\tabularnewline
\end{tabular}
\par\end{centering}
\end{table*}

\subsubsection{\label{sub:Al-Propane}Propane}

\begin{figure}
\begin{centering}
\includegraphics[width=0.5\columnwidth]{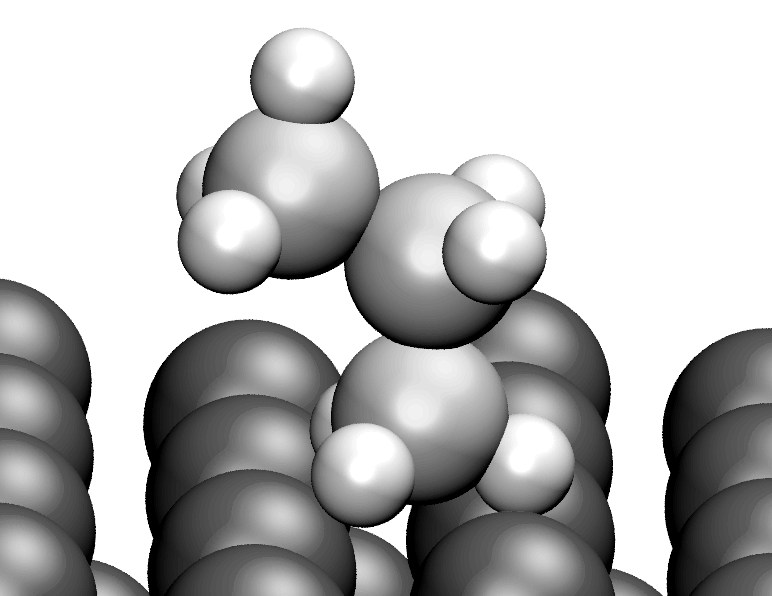}
\par\end{centering}

\caption{\label{fig:Al-propane}A propane molecule on top of the clean Al(111)
surface. Hydrogen atoms are seen as small light gray spheres, carbon
as large light gray spheres, and aluminium as large dark gray spheres.}

\end{figure}

For the case of propane (C$_{3}$H$_{8}$), we tested only the fcc
site on the clean Al(111) surface. The only part of the propane molecule
that can come in contact with the surface, is one of the two CH$_{3}$
groups at the ends. Thus, we tested the case when the propane molecule
is sitting upright on the surface, with only one of the CH$_{3}$
groups touching the surface, as it can be seen in Fig. \ref{fig:Al-propane}.
However, we found no adsorption, rather the propane molecule moved
away from the surface during the relaxation. This was the expected
result, as the propane molecule has no dangling bonds available to
form new bonds to the surface in the CH$_{3}$ endgroup.

\subsubsection{Carbonic acid}

\begin{figure}
\begin{centering}
\includegraphics[width=0.35\columnwidth]{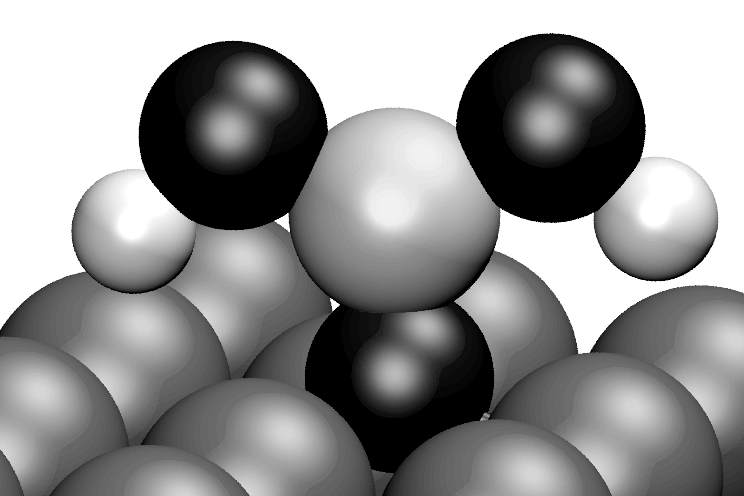}~\includegraphics[width=0.35\columnwidth]{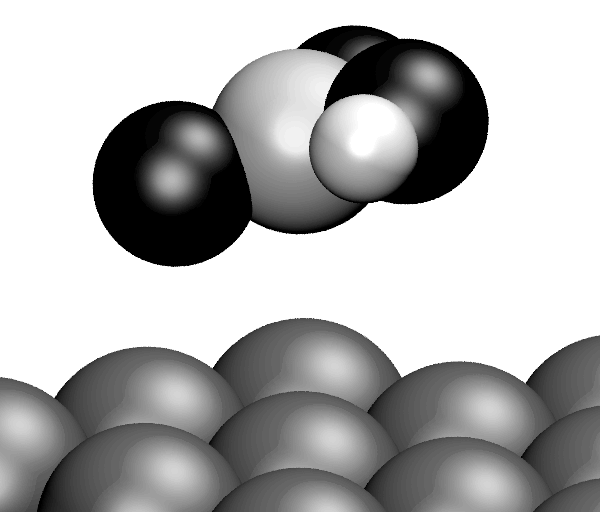}
\par\end{centering}

\caption{\label{fig:al-carbonic-acid}A carbonic acid molecule on top of the
clean Al(111) surface. Start (left) and final configuration (right).
Hydrogen atoms are seen as small light gray spheres, carbon as large
light gray spheres, oxygen as large black spheres, and aluminium as
large dark gray spheres.}

\end{figure}

In Fig. \ref{fig:al-carbonic-acid} one can see a carbonic acid molecule
(H$_{2}$CO$_{3}$) at the fcc site on the clean Al(111) surface.
Due to the hydrogen atoms, only the oxygen atom with the double bond
to the carbon atom is free to react with the surface. Thus, we only
studied the case when the said oxygen was closest to the fcc site,
and no other orientations of the molecule. During the relaxation,
the molecule moved away from the surface, while simultaneously tilting
until lying almost flat in the final configuration. In the final configuration
the molecule was at a distance of $3.4$ Å, and the adsorption energy
was $0.00$ eV, i.e. no adsorption.

\subsection{\label{sub:Al_{2}O_{3}-(0001)-surface}$\alpha$-Al$_{2}$O$_{3}$(0001)
surface}

\begin{figure}
\begin{centering}
\includegraphics[width=0.45\columnwidth]{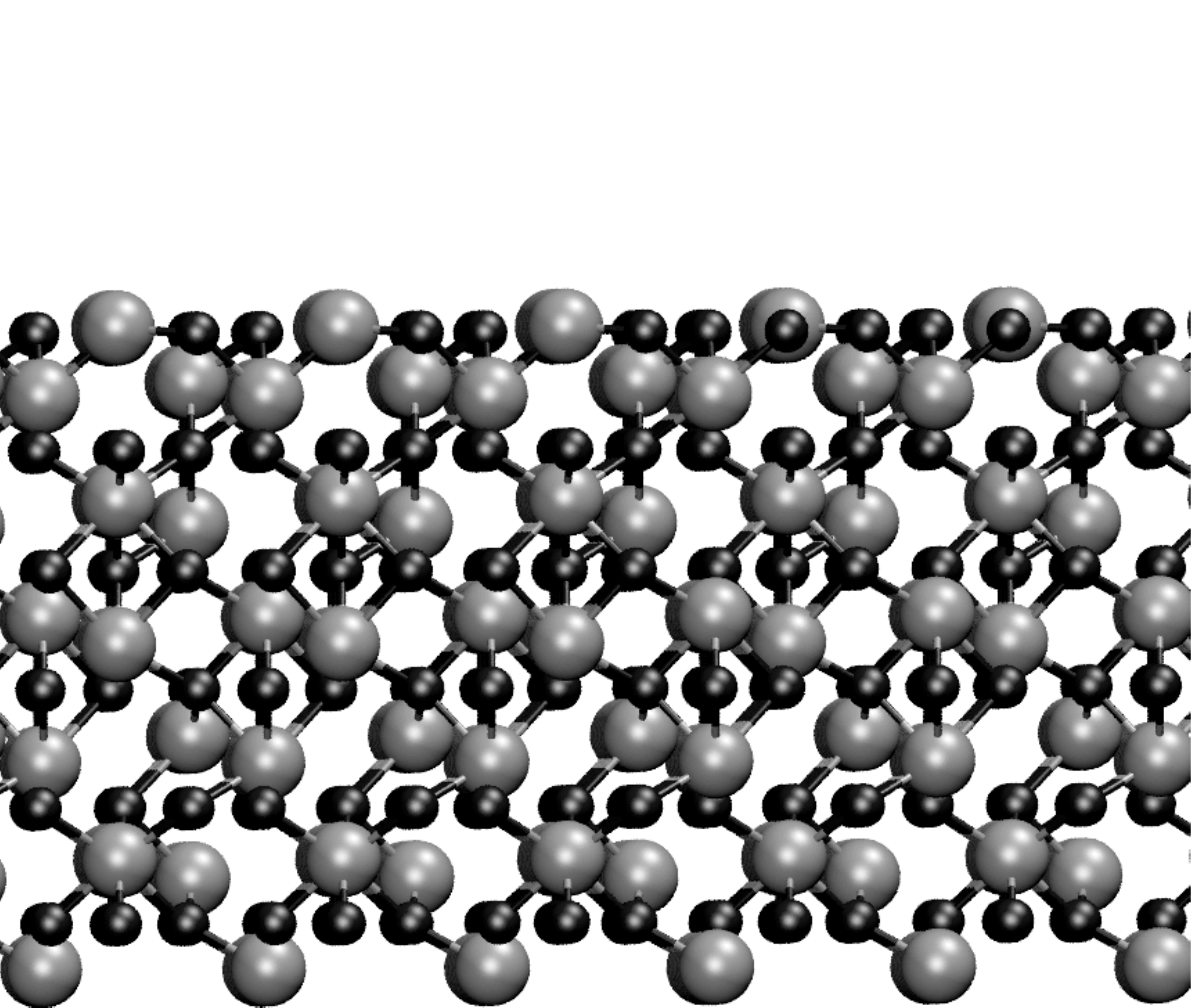}~\includegraphics[width=0.3\columnwidth]{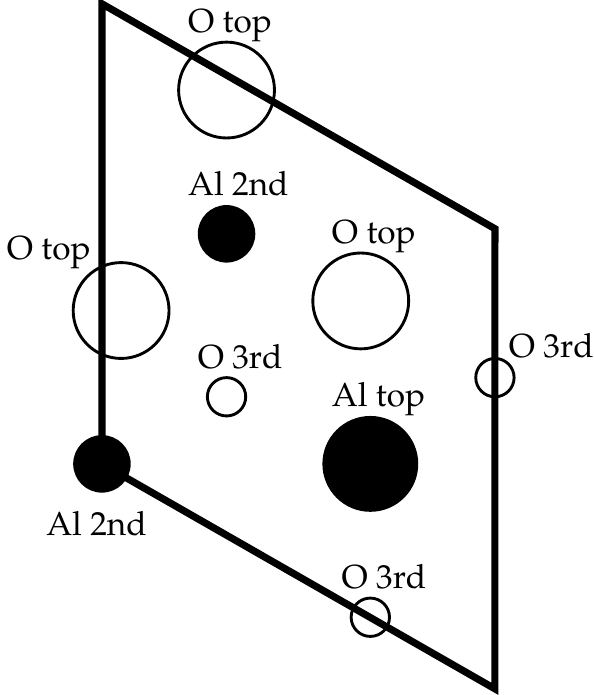}
\par\end{centering}

\caption{\label{fig:Adsorption-sites-Al2O3}General view (left) and surface
atoms (right) on the Al-terminated $\alpha$-Al$_{2}$O$_{3}$(0001)
surface. On the left panel, Al atoms are large light grey spheres,
O atoms small dark grey spheres, and the radius of the spheres is
proportional to the Wigner-Seitz radii of the atoms. On the right,
Al atoms are filled circles, and O atoms empty circles. Due to the
alumina surface relaxation the uppermost Al and O layer are almost
at the same height, then 2nd layer with Al atoms is seen with smaller
circles, and finally the 3rd layer O atoms are seen with still smaller
circles. The primitive surface supercell box can also be seen, whereas
the calculations used a 2x2 surface supercell.}

\end{figure}

A side view of the alumina surface and a schematic top view with adsorption
sites can be seen in Fig. \ref{fig:Adsorption-sites-Al2O3}.

\subsubsection{Benzene}

\begin{figure}
\begin{centering}
\includegraphics[width=0.5\columnwidth]{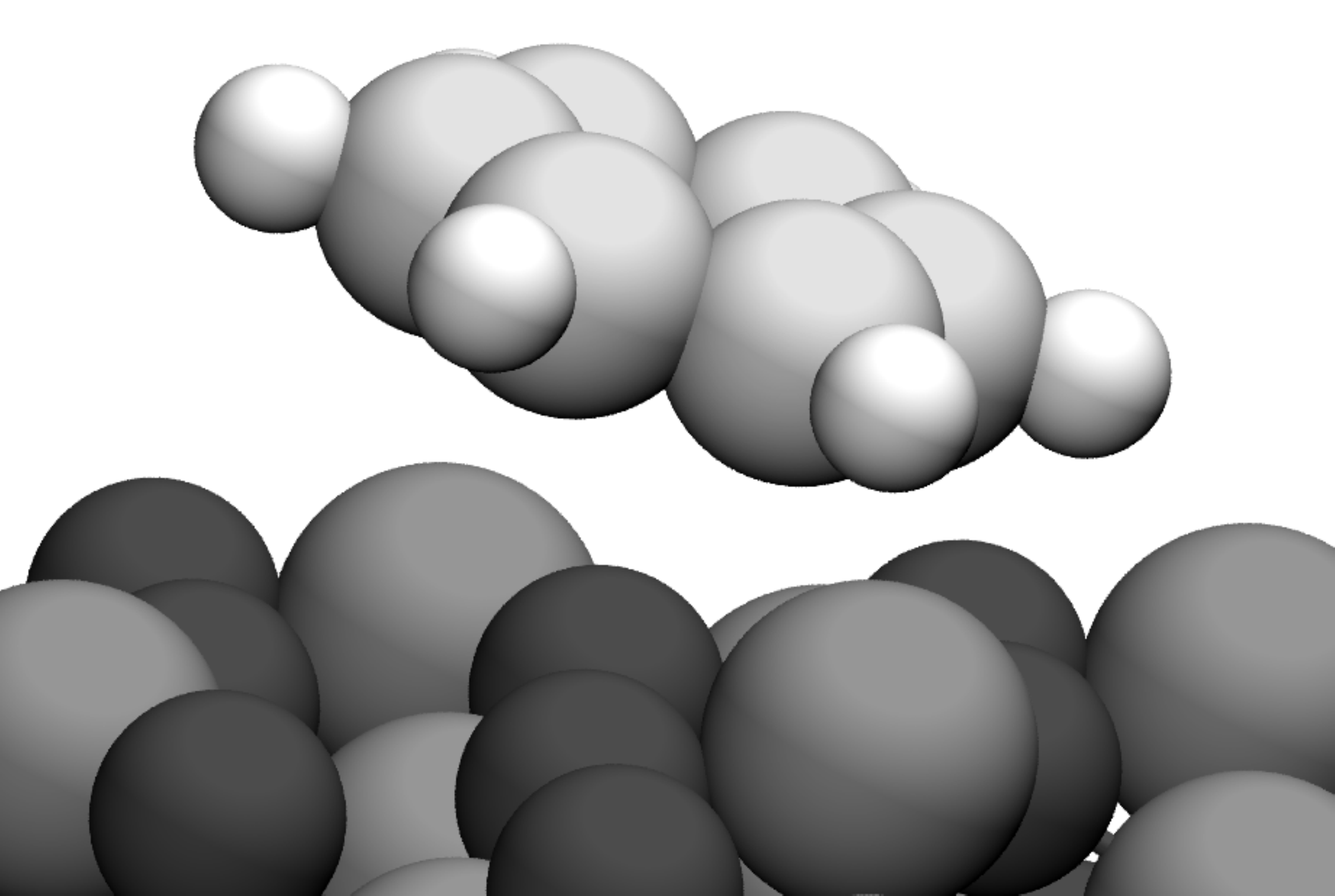}
\par\end{centering}

\caption{\label{cap:alumina-benz-O}A benzene molecule on top of an O atom
on $\alpha$-Al$_{2}$O$_{3}$(0001). Hydrogen atoms are seen as small
light gray spheres, carbon as large light gray spheres, oxygen as
large black spheres, and aluminium as large dark gray spheres.}

\end{figure}

For the alumina surface, the adsorption energies of benzene are also
very weak. A benzene molecule adsorbed at the top site over an O atom
had an adsorption energy of $-0.09$/$-0.05$ (without/with the $\frac{\pi}{6}$
rotation) eV, while on the top site over an Al atom the adsorption
energy was $0.02$/$0.02$ eV. For the hollow size above an Al atom,
the adsorption energy was $0.01$/$0.02$ eV, that is, no adsorption.
For the O top case, where the benzene molecule did not move away into
the vacuum, the molecule preferred to be in a tilted position, with
an angle of $12.8^{\circ}$. This configuration can be seen in Fig.
\ref{cap:alumina-benz-O}. When compared to the case in which the
benzene molecule binds weakly to the Al(111) surface with two nearest
neighbor oxygen atoms on it, we can conclude that increasing further
the number of oxygen atoms in the surface layer decreases the binding
character of C$_{6}$H$_{6}$ on Al(111) surfaces.

\subsubsection{Phenol}

For phenol, we did only a test calculation to check that our adsorption
geometry and energy were in the same ballpark as those reported in
Ref. \cite{chakarova2006}, to which we refer the reader interested
in more detailed calculations for phenol on Al-terminated $\alpha$-Al$_{2}$O$_{3}$(0001).
For the most promising case, labeled (ii) in Ref. \cite{chakarova2006},
we found an adsorption energy of $-0.66$ eV vs. $-1.00$ eV in Ref.
\cite{chakarova2006}. This difference is probably due to different
potentials and parameters used in our simulations. However, the adsorption
geometry was essentially identical, with the phenol molecule adsorbing
at an angle of $42.4^{\circ}$ vs. $44.7^{\circ}$ in Ref. \cite{chakarova2006},
and the OH bond in the phenol molecule bent in a similar way.

\subsubsection{Propane}

For this system two configurations were tested, with the CH$_{3}$
end group of the propane molecule at the Al hollow site and at the
O hollow site on the $\alpha$-Al$_{2}$O$_{3}$(0001) surface. The
configuration of the propane molecule on the surface was the same
as for the Al(111) surface, which can be seen in Fig. \ref{fig:Al-propane}.
In both these cases the propane molecule moved away from the surface
without adsorbing, as also happened with the case of the Al(111) surface.

\subsubsection{Carbonic acid}

\begin{figure}
\begin{centering}
\includegraphics[width=0.5\columnwidth]{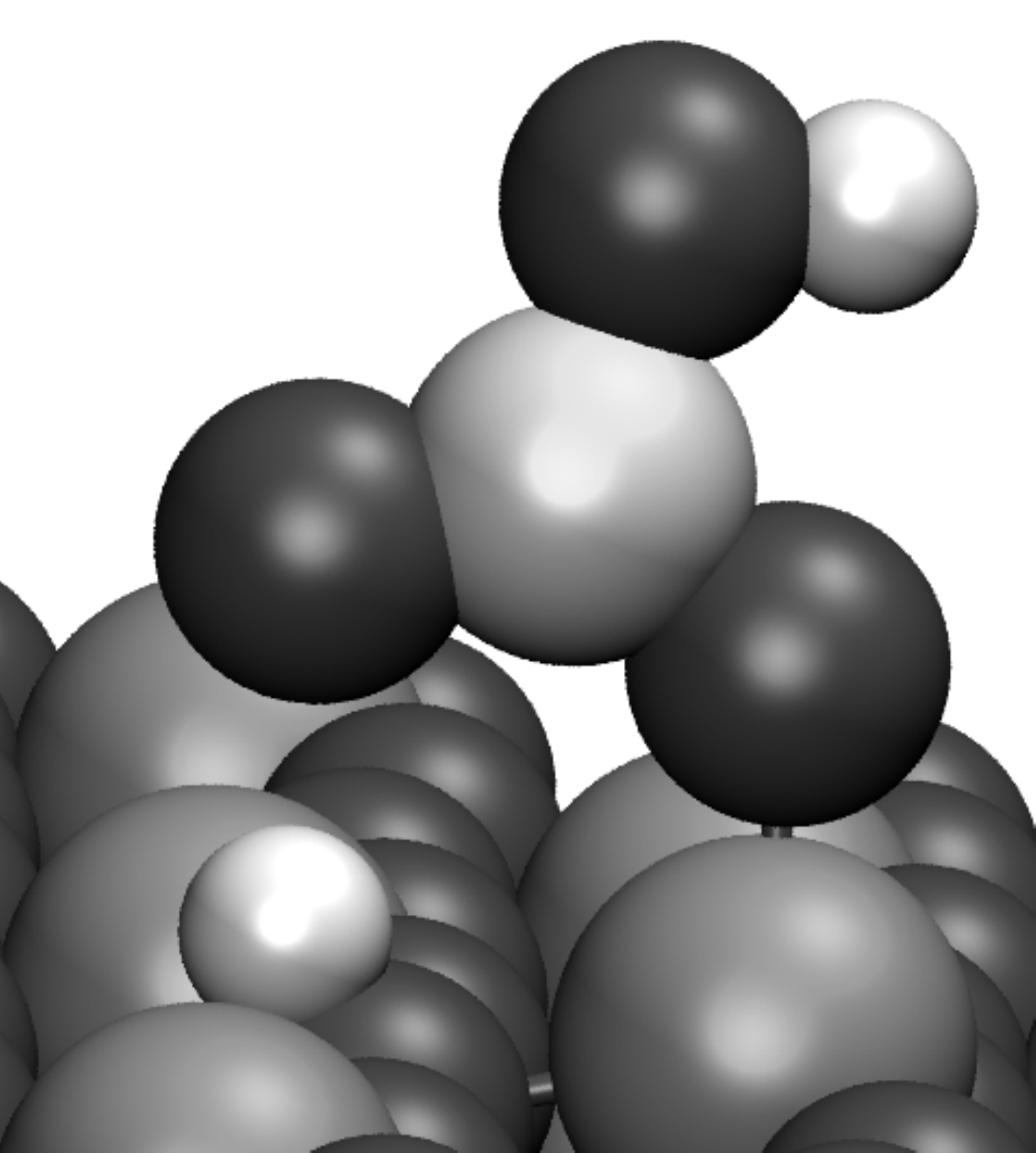}
\par\end{centering}

\caption{\label{fig:alumina_carbonic_acid_side}A carbonic acid molecule adsorbed
on the $\alpha$-Al$_{2}$O$_{3}$(0001) surface. Hydrogen atoms are
seen as small light gray spheres, carbon as large light gray spheres,
oxygen as large black spheres, and aluminium as large dark gray spheres. }

\end{figure}

Contrary to the case of carbonic acid on the Al(111) surface, we found
a significant change in total energy for the carbonic acid molecule
on $\alpha$-Al$_{2}$O$_{3}$(0001). Allowing the carbonic acid molecule
to freely relax on the surface, resulted in the molecule moving to
the Al-top site, with the non-hydrogenated oxygen lying closest to
the surface Al-top atom at a distance of $1.81$ Å. Also, one of the
passivating hydrogens detached from the molecule and adsorbed at the
surface O-top site. This configuration, with the reduction in total
energy of $-1.91$ eV, can be seen in Fig. \ref{fig:alumina_carbonic_acid_side}.
Separating the contributions from the carbonic acid molecule and the
hydrogen atom reveals that much of this energy is due to the rearrangement
of the bonding of the hydrogen atom. The bond dissociation energy
for the hydrogen atom from the carbonic acid molecule in vacuum was
found to be $-5.74$ eV, whereas the adsorption of a hydrogen atom
at the O top site on the $\alpha$-Al$_{2}$O$_{3}$(0001) surface
was $-4.14$ eV. Hence there is a change in the bonding energy of
the hydrogen atom of $+1.60$ eV, and thus in the above system the
contribution of the carbonic acid molecule to the total adsorption
energy of $-1.91$ eV was $-3.50$ eV. Experimental results show qualitateively
similar behaviour \cite{alliot2005,su1997}.

\begin{figure}
\begin{centering}
\includegraphics[width=0.7\columnwidth]{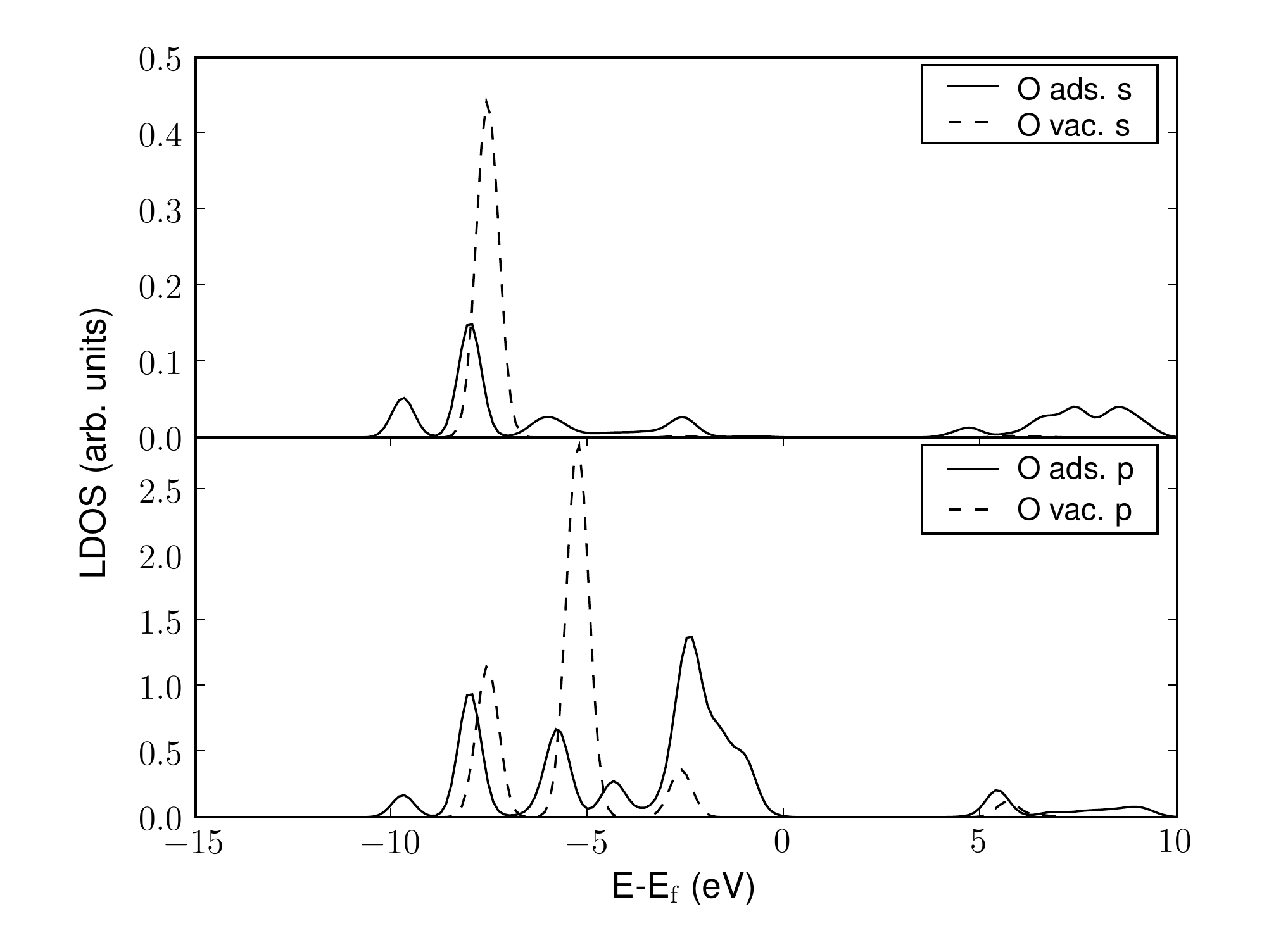}
\par\end{centering}

\caption{\label{fig:alumina_carbonic_acid_O_dos}LDOS of the oxygen atom of
the carbonic acid molecule bound with a double bond (no H passivation
in vacuum) to the carbon atom. In vacuum (dashed lines) and adsorbed
at the Al-top site on the Al$_{2}$O$_{3}$(0001) surface (solid lines).
Both s-LDOS (upper panel) and p-LDOS (lower panel) are shown.}

\end{figure}

Comparing the LDOS for the oxygen atom with the double bond in the
carbonic acid molecule when the molecule is adsorbed at the Al top
site on the $\alpha$-Al$_{2}$O$_{3}$(0001) surface and in vacuum,
we can see what happens when the molecule adsorbs (see Fig. \ref{fig:alumina_carbonic_acid_O_dos}).
One can see that when the molecule is present on the surface, the
LDOS is in general smeared out, and especially for the p-LDOS there
is a shift from the peaks at $-5$ and$-8$ eV towards multiple smeared
out peaks between $-10$ and $0$ eV. Also the s peak around $-8$
eV splits into a band between $-10$ and $-2$ eV. 

\begin{figure}
\begin{centering}
\includegraphics[width=0.7\columnwidth]{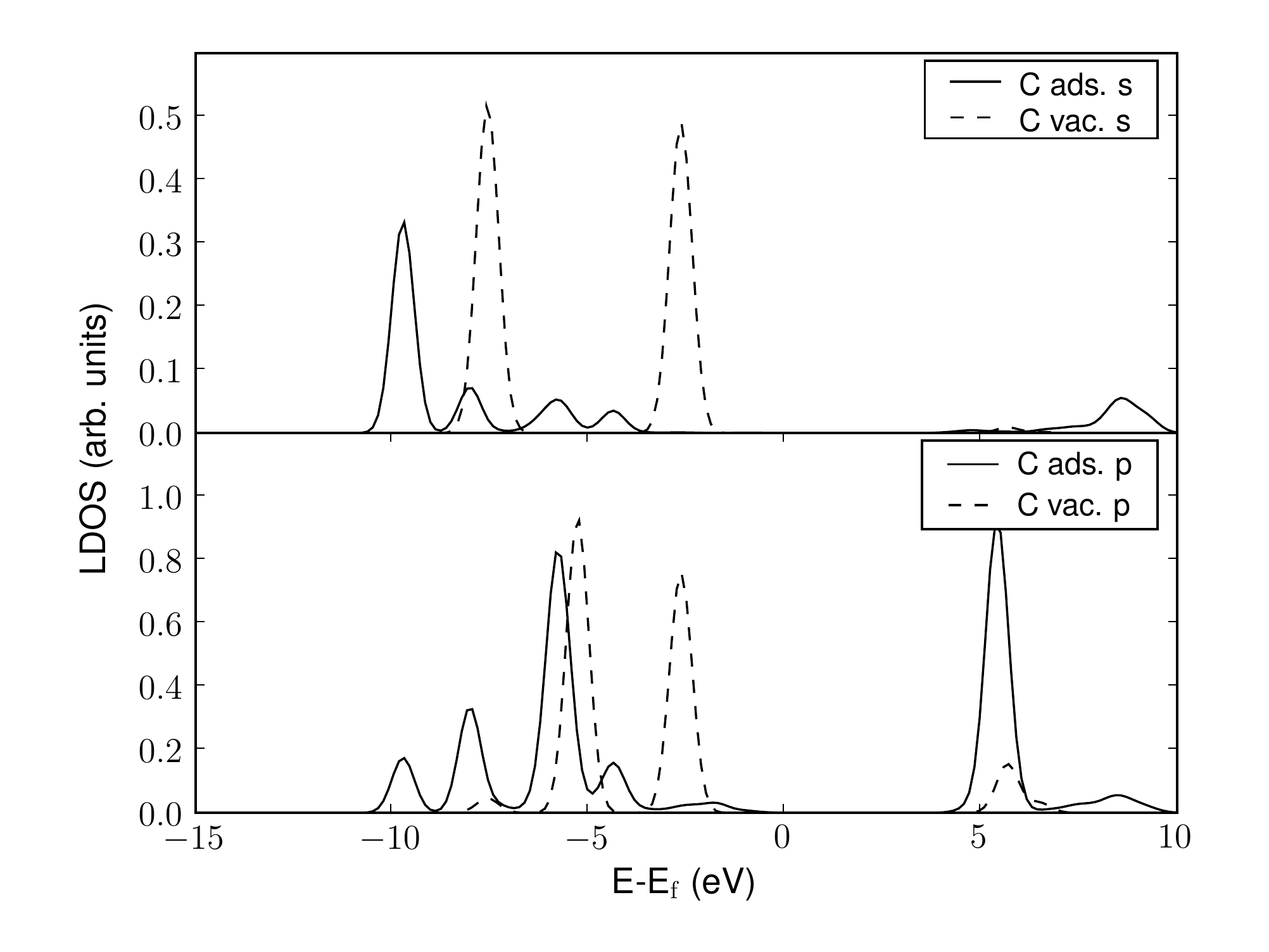}
\par\end{centering}

\caption{\label{fig:alumina_carbonic_acid_C_dos}s-LDOS (upper panel) and p-LDOS
(lower panel) of the carbon atom in the carbonic acid molecule in
vacuum (dashed lines) and adsorbed at the Al-top site on the Al$_{2}$O$_{3}$(0001)
surface (solid lines).}

\end{figure}

\begin{figure}
\begin{centering}
\includegraphics[width=0.7\columnwidth]{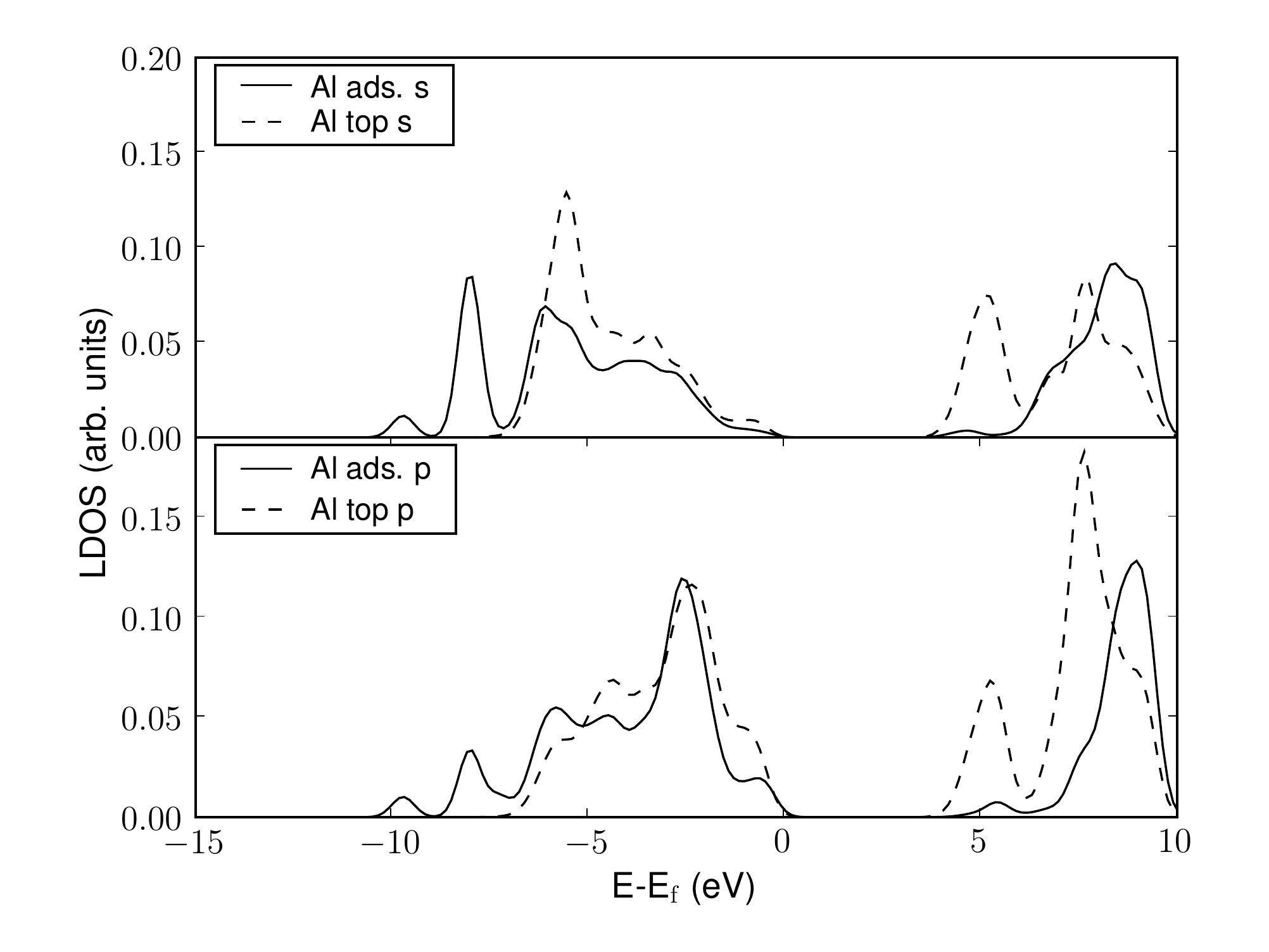}
\par\end{centering}

\caption{\label{fig:alumina_carbonic_acid_Al_dos}s-LDOS (upper panel) and
p-LDOS (lower panel) of the aluminium atom in the surface layer of
the Al$_{2}$O$_{3}$(0001) surface without (dashed lines) and with
(solid lines) the carbonic acid molecule close to the Al atom. }

\end{figure}

Comparing the LDOS's for the oxygen with the LDOS's for the carbon
atom in the carbonic acid molecule (see Figs. \ref{fig:alumina_carbonic_acid_O_dos}
and \ref{fig:alumina_carbonic_acid_C_dos}), one sees similar kind
of broadening of the molecular orbital states of C as for O. On the
other hand, the LDOS's of the aluminium atom on the surface closest
to the carbonic acid molecule also change (see Fig. \ref{fig:alumina_carbonic_acid_Al_dos}).
The s states at $+5$ eV shift to $+8$ eV and the peak at $-6$ eV
below the Fermi level shifts to $-8$ eV. For the p states, the peak
at $+5$ eV vanishes, while the one at $+7$ eV shits up to $+9$
eV. There are also smaller changes in the band between $-10$ and
$0$ eV, showing a shift of the energy bands to lower energies. These
changes match those of the oxygen atom in Fig. \ref{fig:alumina_carbonic_acid_O_dos}.
Thus, the binding character of the alumina surface increases which
can also be seen in the adsorption energy of the carbonic acid molecule
on the surface.


\begin{figure}
\begin{centering}
\includegraphics[width=0.7\columnwidth]{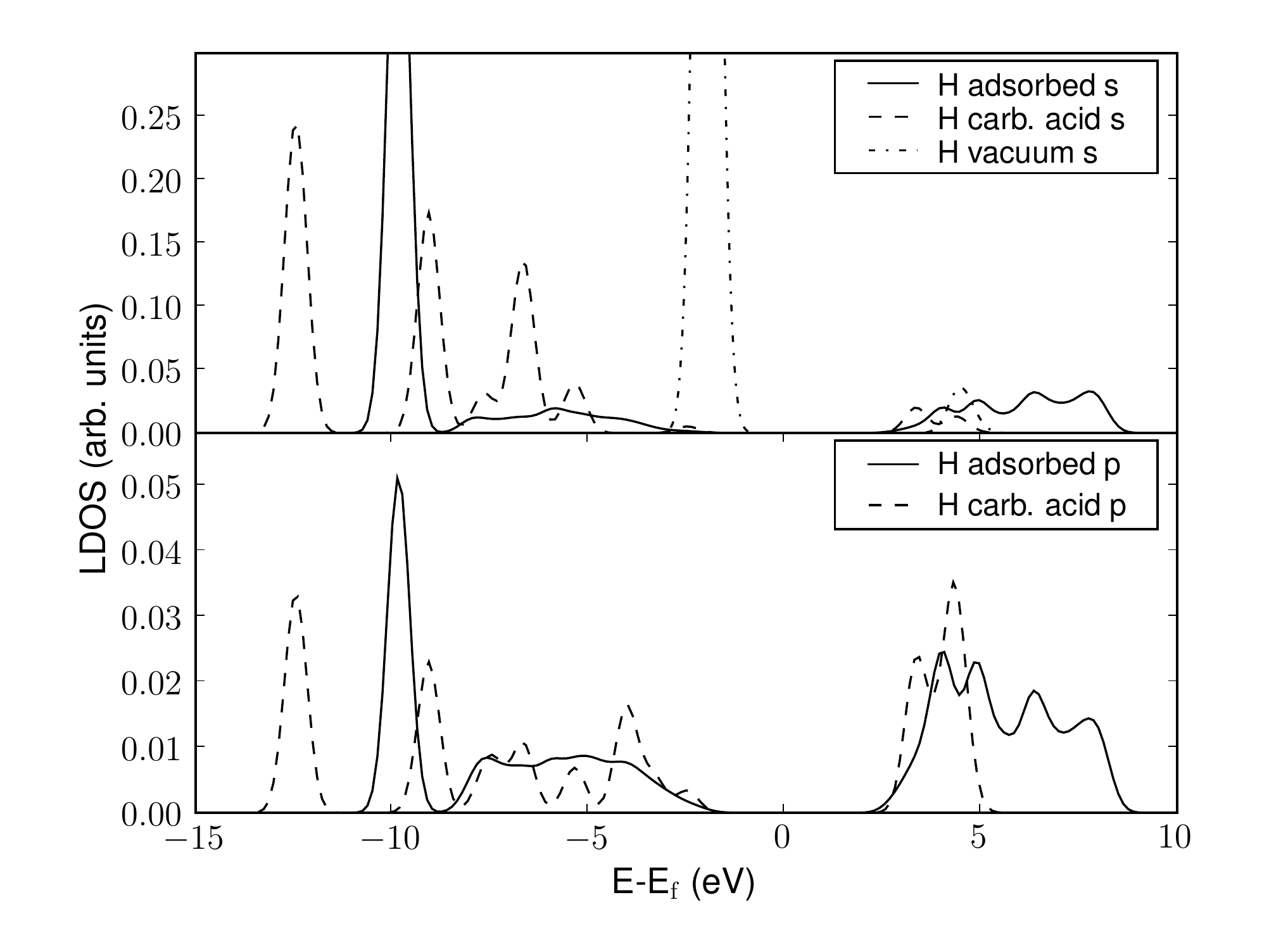}
\par\end{centering}

\caption{\label{fig:alumina_H_DOS}s- and p-LDOS of the H atom at the O-top
site on the alumina surface (solid lines), attached to the carbonic
acid molecule (dashed lines), and in vacuum (chain line).}

\end{figure}

The change in the binding of the H atom as it detaches from the carbonic
acid molecule and attaches to the alumina surface can be seen in Fig.
\ref{fig:alumina_H_DOS}. One can clearly see how the atomic energy
levels split up and hybridize with the molecular orbitals and the
surface band structure, respectively. 

\begin{figure}
\begin{centering}
\includegraphics[width=0.7\columnwidth]{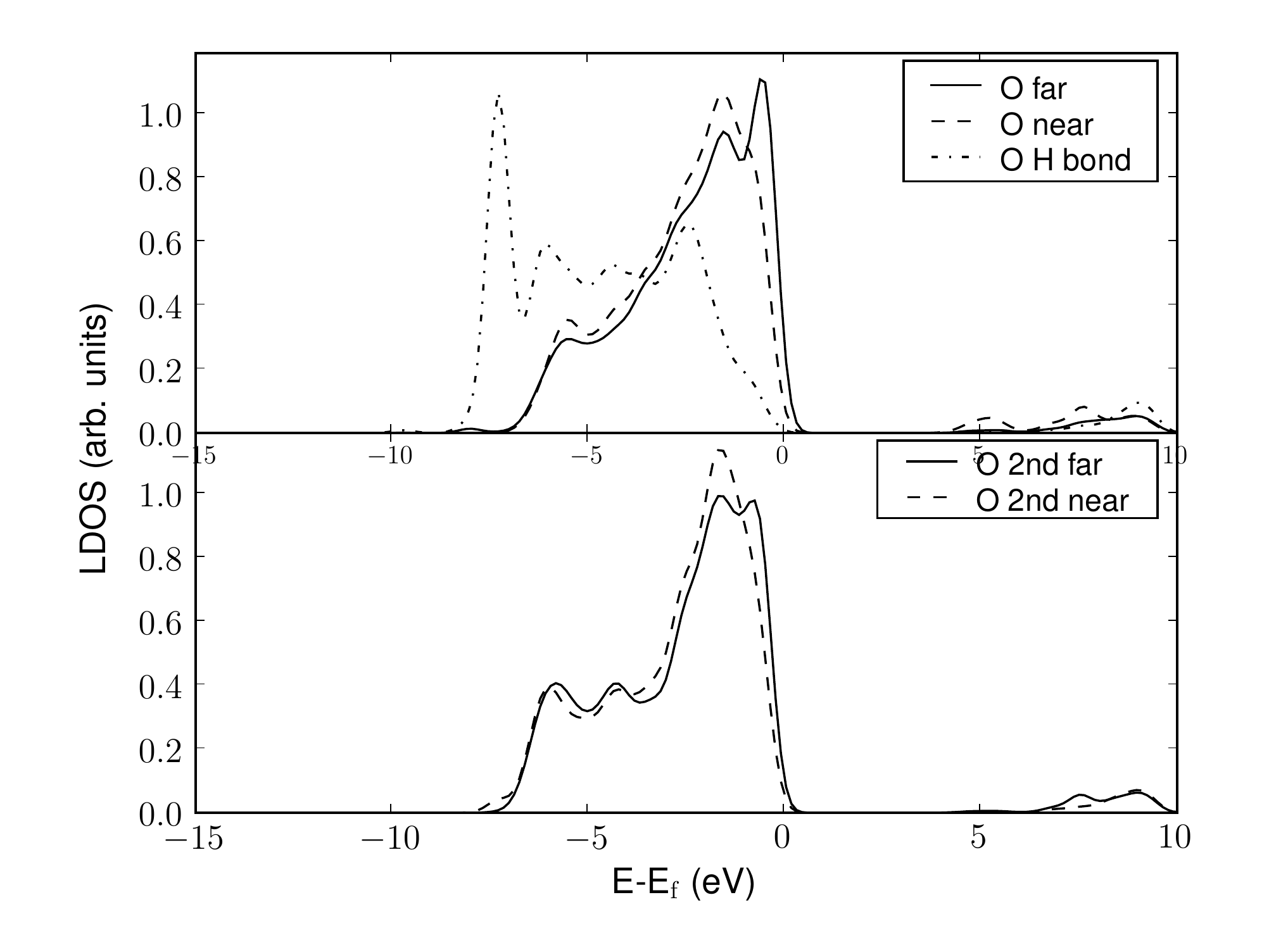}
\par\end{centering}

\caption{\label{fig:alumina_carbonic_acid_surf_O_dos}p-LDOS of the oxygen
atoms in the uppermost (upper panel) and second uppermost (bottom
panel) oxygen layers in the Al$_{2}$O$_{3}$(0001) surface with an
adsorbed carbonic acid molecule. In the upper panel, 'far' refers
to an oxygen atom far away from the adsorbed carbonic acid molecule,
'near' refers to an oxygen atom which is the nearest neighbor of the
Al atom to which the carbonic acid molecule has bound. Finally, 'H
bond' refers to the oxygen atom which the detached hydrogen atom was
bound to. In the bottom panel, '2nd' refers to an oxygen atom in the
second uppermost oxygen layer far away from the carbonic acid molecule,
and '2nd near' refers to an oxygen atom which is the nearest neighbor
of the Al atom to which the carbonic acid molecule has bound.}

\end{figure}

In Fig. \ref{fig:alumina_carbonic_acid_surf_O_dos} one can see the
p-LDOS of the oxygen atoms in the surface layer and in the second
uppermost layer in the surface compared to corresponding oxygen atoms
far away from the carbonic acid adsorption site. For the uppermost
layer, there is a small change in the peak at $-1$ eV, except for
the oxygen atom to which the detached hydrogen atom has bound. In
that case the peak at $-1$eV has disappeared, and instead there is
a new peak at $-10$ eV, along with a wide band between $-8$ and
$-2$ eV which matches the hybridized s and p states of the H atom
in Fig. \ref{fig:alumina_H_DOS}. For the second layer oxygen, there
is a similar shift as for the uppermost layer oxygens, but of a smaller
magnitude.%

\begin{figure}
\begin{centering}
\includegraphics[width=0.35\columnwidth]{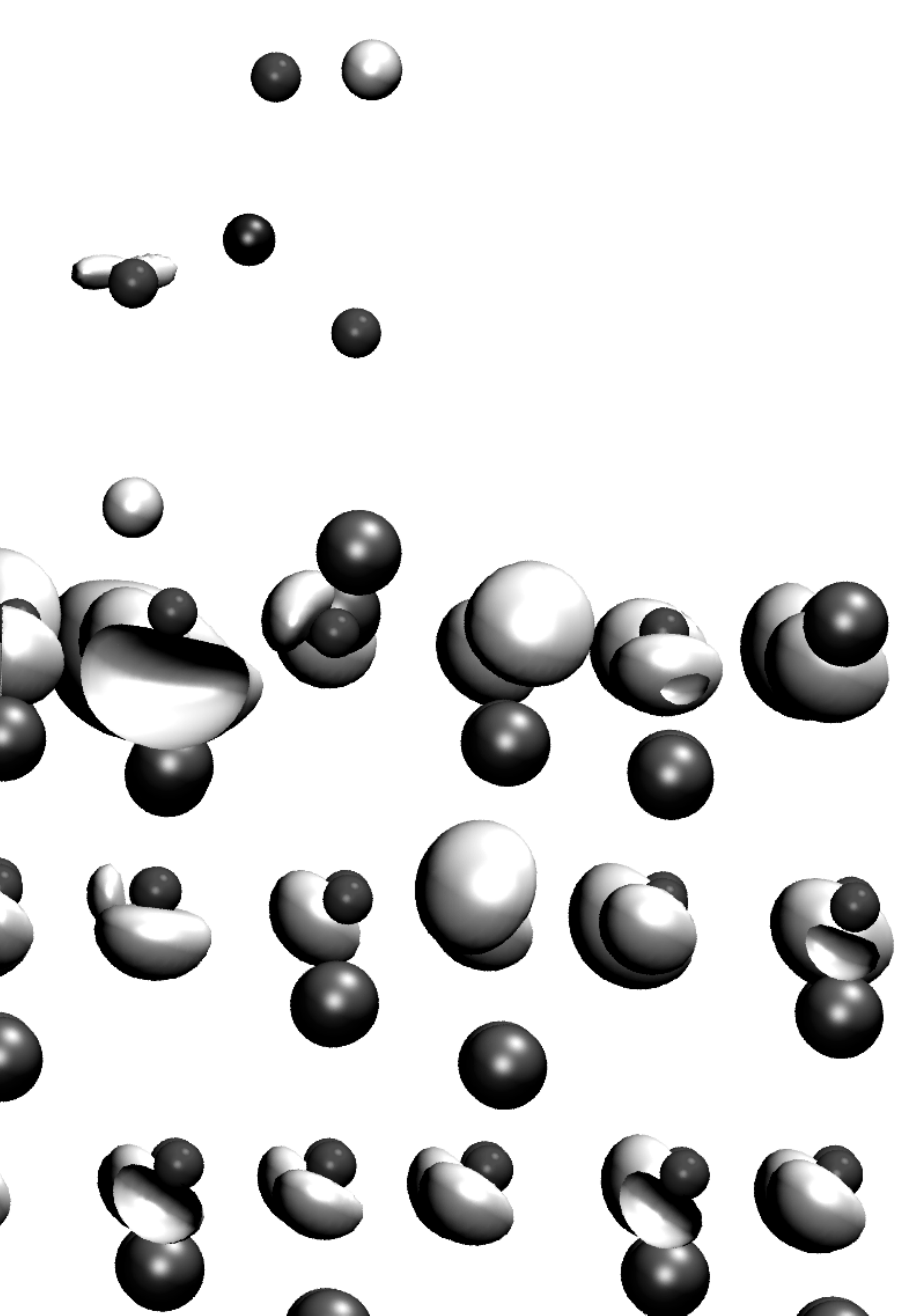}~\includegraphics[width=0.35\columnwidth]{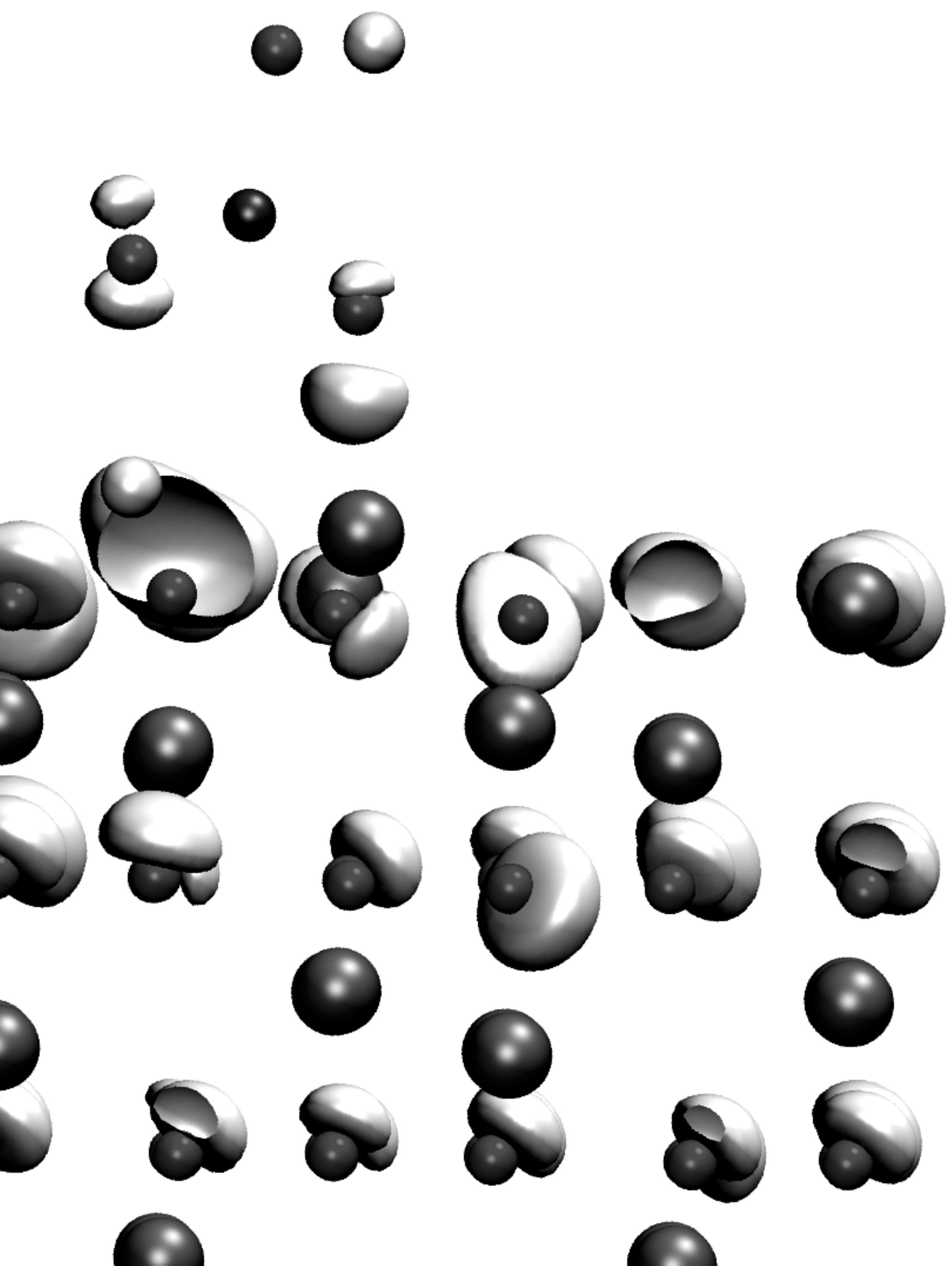}
\par\end{centering}

\caption{\label{fig:Charge-density-difference}Charge density difference for
carbonic acid adsorbed on the alumina surface. On the left, $-0.2$
Å$^{-3}$ isosurface, and on the right $+0.2$ Å$^{-3}$ isosurface.
Al atoms (large gray spheres), O atoms (small gray spheres), C atom
(black sphere), H atoms (white spheres), and isosurfaces (white).}

\end{figure}

Finally, by investigating how the charge density has changed due to
the adsorption (Fig. \ref{fig:Charge-density-difference}), one can
see that the adsorption process causes a large change of charge between
the hydrogen atom on the surface and the closest surface oxygen, as
well as the change of charge between the oxygen closest to the surface
in the carbonic acid molecule and the surface Al atom it is bonded
to. This, like the LDOS analysis above, suggests covalent binding
between both the carbonic acid molecule and the surface, and between
the detached hydrogen atom and the surface. One can also see that
the adsorption causes changes in the charge distribution up to three
oxygen layers down into the substrate, when charge moves upwards to
compensate for the charge that has moved to the adsorption location.

\section{\label{sec:Discussion}Conclusions}

We have shown the results for the adsorption of a few specific small
organic molecules on the oxidized Al(111) and $\alpha$-Al$_{2}$O$_{3}$(0001)
surfaces. In general, the adsorption is quite weak, like for benzene
on oxidized aluminium and phenol on $\alpha$-alumina surfaces, and
the adsorption distance is significant. The major exception is the
carbonic acid molecule which binds strongly to the alumina surface.
In general, as opposed to most transition metals, surface chemistry
on the aluminium surface is dominated by the high affinity of oxygen
to the surface \cite{russell1998}, and thus we expected that the
adsorption for phenol on alumina would be much stronger than for the
benzene molecule. Indeed this was exactly what we saw for carbonic
acid, and to a lesser extent phenol on alumina vs. benzene on alumina.

In the future we plan to test a subset of these results while using
an exchange-correlation functional taking into account van der Waals
(vdW) interactions \cite{dion2004,dion2005}. In Ref. \cite{chakarova2006}
it was found that for phenol on Al-terminated $\alpha$-Al$_{2}$O$_{3}$(0001),
taking vdW into account will enhance adsorption by about $0.2$ eV.
Also, our choice of the RPBE functional, while producing more accurate
covalent molecular adsorption energies, does have the drawback of
not including much of the accidental vdW contribution sometimes seen
with ofter functionals, such as LDA with graphene sheets \cite{hasegawa2004}.
Thus the inclusion of vdW could significantly change the adsorption
energies, especially for the larger molecules like benzene and phenol
where they otherwise adsorbed very weakly or not at all. However,
for the smaller molecules that were strongly covalently bound to the
surface, such as carbonic acid on the alumina surface, the vdW contribution
is expected to be insignificant.

\begin{figure}
\begin{centering}
\includegraphics[width=0.6\columnwidth]{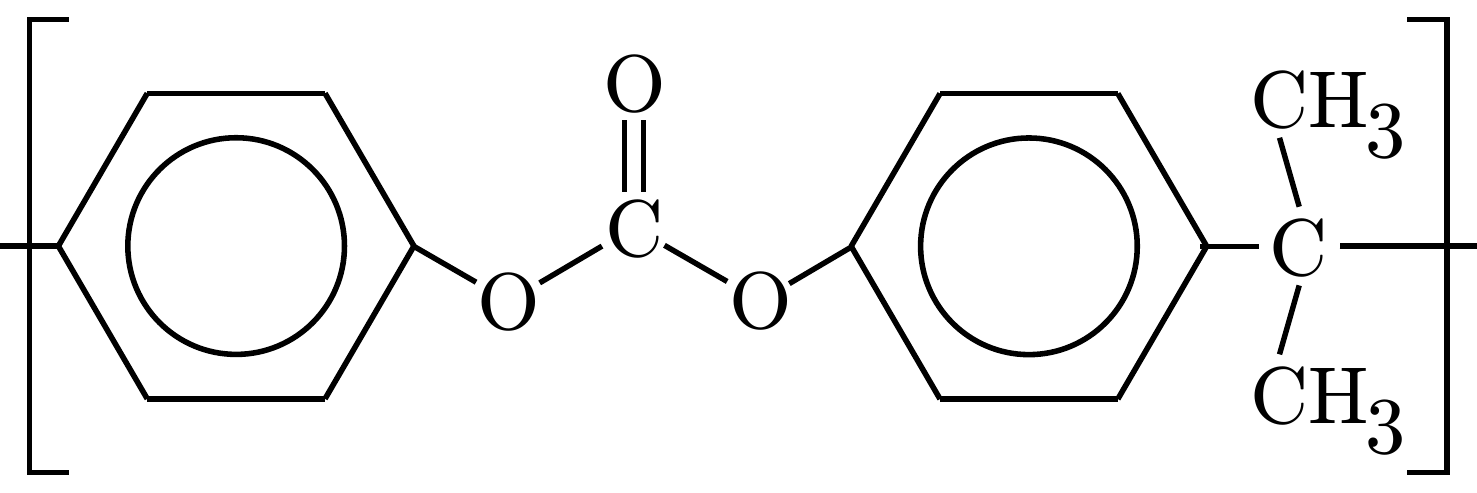}
\par\end{centering}

\caption{\label{fig:BPAPC}A monomer of the bisphenol-A-polycarbonate molecule.
The components from the left are a benzene ring, carbonic acid, another
benzene ring, and finally propane.}

\end{figure}

Moreover, these results allow us to proceed in producing a simplified
potential function for the interaction between BPA-PC and oxidized
Al(111) surfaces as well as BPA-PC and alumina. A schematical figure
of a BPA-PC monomer can be seen in Fig. \ref{fig:BPAPC}. In order
to provide reliable surface interaction energies it is however necessary
to take into account the effect of the nearest neighbor components.
This can be done, e.g. as described in Ref. \cite{schravendijk2007},
such that our current results form the first step in an iterative
procedure. An interesting topic for further study is whether the strong
binding of the carbonic acid plays a relevant role when taking the
neighboring molecules of the BPA-PC monomer into account, as in that
case the passivating hydrogens have been replaced with bonds to the
carbon atoms in neighboring benzene molecules. Also, when the entire
chain is present, the conformations of the chain itself limits the
configurations in which the carbonic acid can come into contact with
the surface. However, for building practical materials from these
substances the pure surface interaction is not necessarily enough
because of the high strains due to the differences in the thermal
expansion coefficients of the metal and the polymer, and one would
need e.g. mechanical adhesion due to pores in the surface, or adhesion
promoters such as glue on the surface. In general it seems that the
details of the surface chemistry do matter a great deal, e.g. for
BPA-PC on Si(001) surfaces hydrogen passivation does completely change
the picture \cite{johnston2007}.

\section*{Acknowledgments}

The authors wish to thank Professor Risto Nieminen at the Helsinki
University of Technology (TKK), Dr. Karen Johnston at the Max Planck
Institute for Polymer Science, and our experimental collaborators
led by Professor Jyrki Vuorinen at the Tampere University of Technology.
This work has been supported by the Finnish Funding Agency for Technology
and Innovation (Tekes) through the HYPRIS project, and by the Academy
of Finland through its Computational Nanoscience (COMP) Center of
Excellence program. We also acknowledge the computer resources provided
by the Finnish IT Center for Science (CSC).

\bibliographystyle{unsrt}
\bibliography{janne}

\begin{thebibliography}{10}

\bibitem{straznicky2004}
P.~V. Straznicky, J.~F. Laliberté, C.~Poon, and A.~Fahr.
\newblock Applications of fiber-metal laminates.
\newblock {\em Polymer Composites}, 21:558--567, Apr 2004.

\bibitem{duschek2000}
R.~Duschek, F.~Mittendorfer, R.~I.~R. Blyth, F.~P. Netzer, J.~Hafner, and M.~G.
  Ramsey.
\newblock The adsorption of aromatics on sp-metals: benzene on {Al(111)}.
\newblock {\em Chemical Physics Letters}, 318:43--48, 2000.

\bibitem{chakarova2006}
Svetla~D. Chakarova-Käck, Øyvind Borck, Elsebeth Schröder, and Bengt~I.
  Lundqvist.
\newblock Adsorption of phenol on graphite(0001) and
  $\alpha$-{Al}$_2${O}$_3$(0001): Nature of van der {W}aals bonds from
  first-principles calculations.
\newblock {\em Physical Review B}, 74:155402, October 2006.

\bibitem{su1997}
Chunming Su and Donald~L. Suarez.
\newblock In sity infrared speciation of adsorbed carbonate on aluminum and
  iron oxides.
\newblock {\em Clays and Clay Minerals}, 45(6):814--825, 1997.

\bibitem{alliot2005}
C.~Alliot, L.~Bion, F.~Mercier, and P.~Toulhoat.
\newblock Sorption of aqueous carbonic, acetic, and oxalic acids onto
  $\alpha$-alumina.
\newblock {\em Journal of Colloid and Interface Science}, 287(2):444--451, July
  2005.

\bibitem{johnston2007}
Karen Johnston and Risto~M. Nieminen.
\newblock Polymer adhesion: First-principles calculations of the adsorption of
  organic molecules onto {Si} surfaces.
\newblock {\em Physical Review B}, 76:085402, 2007.

\bibitem{kresse1993}
Georg Kresse and J.~Hafner.
\newblock Ab initio molecular dynamics for liquid metals.
\newblock {\em Physical Review B}, 47(1):558--561, January 1993.

\bibitem{kresse1996a}
Georg Kresse and Jürgen Furthmüller.
\newblock Efficiency of ab-initio total energy calculations for metals and
  semiconductors using a plane-wave basis set.
\newblock {\em Computational Materials Science}, 6:15--50, July 1996.

\bibitem{kresse1996b}
Georg Kresse and Jürgen Furthmüller.
\newblock Efficient iterative schemes for ab initio total-energy calculations
  using a plane-wave basis set.
\newblock {\em Physical Review B}, 54(16):11169--11186, October 1996.

\bibitem{hammer1999}
Björk Hammer, L.B. Hansen, and J.K. Nørskov.
\newblock Improved adsorption energetics within density functional theory using
  revised {P}erdew-{B}urke-{E}rnzerhof functionals.
\newblock {\em Physical Review B}, 59:7413--7421, Mar 1999.

\bibitem{perdew1996}
John~P. Perdew, Kieron Burke, and Matthias Ernzerhof.
\newblock Generalized gradient approximation made simple.
\newblock {\em Physical Review Letters}, 77:3865, 1996.

\bibitem{blochl1994}
P.~E. Blöchl.
\newblock Projector augmented-wave method.
\newblock {\em Physical Review B}, 50(24):17953, 1994.

\bibitem{kresse1999}
G.~Kresse and D.~Joubert.
\newblock From ultrasoft pesudopotentials to the projector augmented-wave
  method.
\newblock {\em Physical Review B}, 59(3):1758--1775, Jan 1999.

\bibitem{godin1994}
T.~J. Godin and John~P. Lafemina.
\newblock Atomic and electronic structure of the corundum ($\alpha$-alumina)
  (0001) surface.
\newblock {\em Physical Review B}, 49(11):7691, March 1994.

\bibitem{monkhorst1976}
Hendrik~J. Monkhorst and James~D. Pack.
\newblock Special points for {B}rilliouin-zone integrations.
\newblock {\em Physical Review B}, 13(12):5188, 1976.

\bibitem{pack1977}
James~D. Pack and Hendrik~J. Monkhorst.
\newblock Special points for {B}rillouin-zone integrations - a reply.
\newblock {\em Physical Review B}, 16(4):1748, August 1977.

\bibitem{methfessel1989}
M.~Methfessel and A.~T. Paxton.
\newblock High-precision sampling for brillouin-zone integration in metals.
\newblock {\em Physical Review B}, 40(6):3616--3621, August 1989.

\bibitem{CRC86}
David~R. Lide, editor.
\newblock {\em CRC Handbook of Chemistry and Physics}.
\newblock CRC Press, 86th edition, 2005.

\bibitem{perdew1992}
John~P. Perdew, J.A. Chevary, S.H. Vosko, K.~A. Jackson, M.~R. Pederson, D.J.
  Singh, and C.~Fiolhais.
\newblock Atoms, molecules, solids, and surfaces: Applications of the
  generalized gradient approximation for exchange and correlation.
\newblock {\em Physical Review B}, 46(11):6671, 1992.

\bibitem{mattsson2006}
Ann~E. Mattsson, Rickard Armiento, Peter~A. Schultz, and Thomas~R. Mattsson.
\newblock Nonequivalence of the generalized gradient approximations {PBE} and
  {PW91}.
\newblock {\em Physical Review B}, 73:195123, 2006.

\bibitem{yourdshahayan2002}
Y.~Yourdshahayan, B.~Razaznejad, and B.~I. Lundqvist.
\newblock Adiabatic potential-energy surfaces for oxygen on {Al(111)}.
\newblock {\em Physical Review B}, 65:075416, 2002.

\bibitem{kiejna2001}
A.~Kiejna and B.~I. Lundqvist.
\newblock First-principles study of surface and subsurface {O} structures at
  {Al(111)}.
\newblock {\em Physical Review B}, 63:085405, 2001.
\newblock (64) 049901(E).

\bibitem{kiejna2002}
A.~Kiejna and B.~I. Lundqvist.
\newblock Stability of oxygen adsorption sites and ultrathin aluminium oxide
  films on {Al(111)}.
\newblock {\em Surface Science}, 504:1--10, 2002.

\bibitem{russell1998}
Jr. J.~N.~Russel, A.~Leming, and R.~E. Morris.
\newblock Phenol decomposition on {Al(111)}.
\newblock {\em Surface Science}, 399:239--247, 1998.

\bibitem{dion2004}
M.~Dion, H.~Rydberg, E.~Schröder, D.~C. Langreth, and B.~I. Lundqvist.
\newblock Van der {W}aals density functional for general geometries.
\newblock {\em Physical Review Letters}, 92:246401, June 2004.

\bibitem{dion2005}
M.~Dion, H.~Rydberg, E.~Schröder, D.~C. Langreth, and B.~I. Lundqvist.
\newblock Erratum: Van der {W}aals density functional for general geometries
  [{P}hys. {R}ev. {L}ett. 92, 246401 (2004)].
\newblock {\em Physical Review Letters}, 95:109902, September 2005.

\bibitem{hasegawa2004}
Masayuki Hasegawa and Kazume Nishidate.
\newblock Semiempirical approach to the energetics of interlayer binding in
  graphite.
\newblock {\em Physical Review B}, 70(20):205431+, November 2004.

\bibitem{schravendijk2007}
Pim Schravendijk, Luca~M. Ghiringhelli, Luigi~Delle Site, and Nico~F.A. van~der
  Vegt.
\newblock Interaction of hydrated amino acids with metal surfaces: A multiscale
  modeling description.
\newblock {\em Journal of Physical Chemistry C}, 111:2631--2642, 2007.

\end{thebibliography}

\end{document}